%% file: talk.tex
\documentclass[12pt]{article}
\usepackage{epsfig}

\input econfmacros.tex
\newcommand{\bea}{\begin{eqnarray}}
\newcommand{\eea}{\end{eqnarray}}
\newcommand{\bay}{\begin{array}}
\newcommand{\eay}{\end{array}}

\def\Title#1{\begin{center} {\Large {\bf #1} } \end{center}}

\begin{document}
\begin{flushright}
UCSD/PTH 02-16\\hep-ph/0207095
\end{flushright}

\Title{Probing New Physics with $b\to s\gamma$ decays\footnote{Talk given at 
{\em Flavor Physics and CP Violation (FPCP)}, Philadelphia, PA, May 2002.}}

\bigskip\bigskip


\begin{raggedright}  

{\it Dan Pirjol\index{Pirjol, D.}\\
Department of Physics, UCSD\\
9500 Gilman Drive\\
La Jolla, CA 92093}
\bigskip\bigskip
\end{raggedright}

\section*{Abstract}

In the Standard Model,
the photon emitted in $b\to s\gamma$ decays is predicted to be left-handed 
polarized. 
We discuss the types of New Physics which can produce a deviation from
this prediction, focusing on the Minimal Supersymmetric Standard Model.
A new method is proposed for testing these predictions, which makes use of
angular correlations in exclusive $B\to K_{\rm res}(\to K\pi\pi)\gamma$ decays.
\vspace{0.5cm}


Rare radiative $b\to s\gamma$ decays have been extensively investigated
both as a probe of the flavor structure of the Standard Model and for
their sensitivity to any new physics beyond the SM. 
The present experimental average of the inclusive rate
$Br(B\to X_s\gamma) = (3.22\pm 0.40)\times 10^{-4}$  \cite{data}
agrees well with the Standard Model prediction. To next-to-leading 
order in perturbation theory one obtains
$Br(B\to X_s\gamma) = (3.54\pm 0.49)\times 10^{-4}$
(corresponding to a pole mass ratio $m_c/m_b = 0.25\pm 0.06$) \cite{AGL}. 
In addition to the rather well predicted inclusive branching ratio,
there is a unique feature of this process within the SM which drew only 
moderate theoretical attention and which has not yet been tested.
Namely, the emitted photons are left-handed in radiative $B^-$ and $\bar B^0$
decays and are right-handed in $B^+$ and $B^0$ decays. In the SM the photon in
$b \to s\gamma$ is predominantly left-handed, since only left chiral 
quarks couple to the $W$ into loops. 

This prediction holds in the SM to within a few percent, up   
to corrections of order $\Lambda/m_b$, for exclusive and 
inclusive decays. On the other hand, in certain extensions of the Standard
Model, an appreciable right-handed component can be induced in $b\to s\gamma$
decays. While measurements 
of the inclusive radiative decay rate agree with SM calculations,
no evidence exists so far for the helicity of the photons in these decays.

In view of its popularity as a model of New Physics, we will focus the discussion
below on the Minimal Supersymmetric Standard Model (MSSM) \cite{MSSM} 
(an alternative source of nonstandard photon helicity is the left-right 
symmetric model \cite{LRM}). In addition
to the $t-W^\pm$ penguin loop, which is responsible for the $b\to s\gamma$ 
decay in the Standard Model, the MSSM allows for new contributions. These
come from loops containing at least one charged particle, and include
the top-charged Higgs contribution $t-H^\pm$, stop-chargino
$\tilde t-\tilde \chi^\pm$ and sbottom-neutralino/gluino contributions $\tilde
\chi^0-\tilde b$, $\tilde g-\tilde b$.

The flavor structure of the MSSM is not determined by symmetries
and is generated by the Yukawa terms in the superpotential and the
soft SUSY breaking terms which introduce many arbitrary parameters, 
and in general can lead to FCNC transitions
(SUSY flavor problem). They are naturally constrained in models with minimal
flavor violation (MFV), where the only source of flavor violation is the usual
CKM matrix. The most general form of the squark mass matrices in
the super-CKM basis can be written as (see, e.g., \cite{mass})
\bea
{\cal M}_U^2 = \left(
\begin{array}{cc}
M_{U_{LL}}^2 & M_{U_{LR}}^2 \\
M_{U_{LR}}^{2\dagger} & M_{U_{RR}}^2 \\
\end{array}
\right)\,,\qquad
{\cal M}_D^2 = \left(
\begin{array}{cc}
M_{D_{LL}}^2 & M_{D_{LR}}^2 \\
M_{D_{LR}}^{2\dagger} & M_{D_{RR}}^2 \\
\end{array}
\right)
\eea
The $3\times 3$ submatrices $M_{U_{ij}}^2$ and $M_{D_{ij}}^2$ are given in
terms of the quark mass matrices and soft SUSY breaking terms 
$M_{\tilde U_{L,R}}^2$, $M_{\tilde D_{L,R}}^2$ and $A_U, A_D$ defined as usual
by
\bea
{\cal L}_{soft} = -\sum_{\tilde Q=\tilde U_L,\tilde D_L}
\tilde Q^\dagger M_{\tilde Q_L}^2 \tilde Q - 
\tilde U^\dagger M_{\tilde U_R}^2 \tilde U - 
\tilde D^\dagger M_{\tilde D_R}^2 \tilde D + \tilde Q A_U H_U \tilde U
+ \tilde Q A_D H_D \tilde D\,.
\eea
In MFV models the matrices $A_U, A_D$, $M_{\tilde U_{R}}^2$ and 
$M_{\tilde D_{R}}^2$ must be diagonal. Usually, this is taken to imply
that the only contribution to $b\to s\gamma$ in MFV-MSSM is the
top-charged Higgs diagram, which has the same dominant chiral structure as the
SM. In such a situation, the photon in $b\to s\gamma$ is again left-handed.

It was pointed out in \cite{BEFU} that MFV models actually allow nontrivial
flavor violation in the squark sector. The matrices $M_{\tilde U_{L}}^2$
and $M_{\tilde D_{L}}^2$ are connected by SU$_{\rm L}$(2) gauge invariance as
$M_{\tilde D_{L}}^2 = V_{CKM}^\dagger M_{\tilde U_{L}}^2 V_{CKM}$, which 
implies that a diagonal, but not proportional to the unit matrix 
$M_{\tilde D_{L}}^2$, can give a non-diagonal structure for $M_{\tilde U_{L}}^2$
(and vice versa). This allows the chargino-up squark and neutralino/gluino-down
squark contributions to $b\to s\gamma$. The latter graphs can occur with a
helicity flip along the gluino line, which can produce a right-handed photon
component in $b\to s\gamma$.

Relaxing the MFV constraints on the flavor structure (the so-called
unconstrained MSSM) generally leaves the new physics contributions to
$b\to s\gamma$ be dominated by the gluino graph, which can easily introduce
a right-handed photon component. Data on the total $B\to X_s\gamma$ branching 
ratio 
set very stringent constraints on the allowed squark mass matrices
\cite{generic}. An extreme way of satisfying such constraints in a generic 
MSSM has been proposed in \cite{CR}, where it is suggested that the MSSM
graphs exactly cancel the SM contribution to the left-handed penguin amplitude,
in such a way that the right-handed amplitude precisely reproduces the 
observed rate. A measurement of the photon helicity in $b\to s\gamma$ will 
clearly help decide which of these possibilities (if any) is realized in Nature.

Several ways were suggested to look for signals of 
physics beyond the SM through photon helicity effects in $B \to X_s \gamma$.
In  the first suggested method \cite{AGS} the photon helicity is 
probed through mixing-induced CP asymmetries. The sensitivity to the polarization
comes from interference between $B^0$ and $\bar B^0$ decay amplitudes into a 
common state of definite photon polarization. However, measuring asymmetries 
at a level of a few percent, as expected in the SM, require an order of $10^9$ 
$B$ mesons which might not be available at the existing B factories for some 
time. In a second scheme one studies angular distributions in
$B\to \gamma (\to e^+e^-) K^* (\to K\pi)$, where the photon can be 
virtual \cite{Kim} or real, converting in the beam pipe to an electron-positron 
pair \cite{GrPi}. The efficiency of this method is comparable to that of the
previous method. A somewhat different method was proposed in \cite{Lambda} and 
makes use of angular correlations in both exclusive and inclusive 
$\Lambda_b\to X_s\gamma$ decays.

We discuss in the following a method \cite{IsWe,GGPR,GroPi} for measuring the 
photon 
polarization using angular correlations in the strong decay of 
a $K_{\rm res}$ resonance produced in $B\to K_{\rm res}\gamma$. 
Denoting with $A_{R,L}(\vec p_i)$ the amplitude for the strong decay
of a $K_{\rm res}$ at rest in a spin state $|j,m=\pm 1\rangle$ into a final
state $|f, \vec p_i\rangle$ containing spinless hadrons with momenta 
$\vec p_i$, one could
ask what is the condition for a nonvanishing asymmetry $|A_R(\vec p_i)|\neq 
|A_L(\vec p_i)|$. This is a typical 'motion-reversal' asymmetry of the form
\bea\label{Todd}
a_{i\to f}^{T-odd} \equiv |T_{i\to f}|^2 - |T_{\bar i\to \bar f}|^2
\eea
where $\bar i$, $\bar f$ are motion-reversed states, obtained by changing the 
momenta
and spins $(\vec p_i, s_i)\to (-\vec p_i, -s_i)$. 
Since parity is conserved in strong interactions, one can replace
$|T_{\bar i\to \bar f}| = |T_{P\bar i\to P\bar f}|$ on the right-hand 
side to recover the polarization asymmetry $|A_R(\vec p_i)|^2-|A_L(\vec p_i)|^2$.
$T$-invariance (or equivalently CP invariance) of the strong interactions
gives $|T_{\bar i\to \bar f}| = |T_{f\to i}|$. Using this together with 
the unitarity condition $T_{i\to f}^* - T_{f\to i} =
-i\sum_k T_{i\to k}^* T_{f\to  k} \equiv -i\alpha_{i\to f}$ into (\ref{Todd})
gives that the left/right asymmetry can be written as
\bea
a_{i\to f}^{T-odd} = 2\mbox{Im }(T_{i\to f} \alpha_{i\to f}) - 
|\alpha_{i\to f}|^2\,.
\eea
If all decay amplitudes are real then $\alpha_{i\to f}=0$ which shows that
a nonvanishing asymmetry requires nontrivial final state interactions.

Furthermore, 2-body final states (e.g. $K\pi$) cannot produce an asymmetry 
because it is impossible to form a quantity which is odd under motion reversal
from just two vectors $\vec q$ (photon momentum in the $K_{\rm res}$ frame) 
and $\hat n$ (the direction parameterizing the final state 
$|K(\hat n)\pi(-\hat n)\rangle$. 

A nonvanishing asymmetry can be realized however in 3-body strong decays 
$K_{\rm res}\to K\pi\pi$. Such decays are realized for the lowest excitations
of the $K$ with quantum numbers $J^P = 1^-, 1^+, 2^+$, some of which have
been recently observed to be produced in rare radiative decays. Both Belle
and CLEO measured recently the decay $B\to K_2^*(1430)\gamma$ with a branching 
ratio of $(1.50^{+0.58+0.11}_{-0.53-0.13})\times 10^{-5}$ and
$(1.66^{+0.59}_{-0.53}\pm 0.13)\times 10^{-5}$, respectively \cite{exp}.
Similar branching ratios are expected from theoretical estimates for
decays into $K_1(1400)$ and $K_1(1270)$ \cite{theory}.

These states decay strongly to 3-body final $K\pi\pi$ states. Neglecting
a small nonresonant contribution, these decays are dominated by interference
of a few channels
\bea\label{chain}
K_{\rm res}^{+}\to 
\left\{
\begin{array}{c}
 K^{*+}\pi^0 \\
 K^{*0} \pi^+ \\
 \rho^+ K^0
\end{array}
\right\} \to K^0 \pi^+ \pi^0\,,\qquad
K_{\rm res}^0\to 
\left\{
\begin{array}{c}
 K^{*+}\pi^- \\
 K^{*0} \pi^0 \\
 \rho^- K^+
\end{array}
\right\} \to K^+ \pi^- \pi^0\,.
\eea
The different channels $K^*\pi$ are related by isospin symmetry and
contribute with a relative strong phase which can be parameterized
in terms of Breit-Wigner forms. The $K_1(1400)$ decays predominantly
to  $K^*\pi$ in a mixture of $S$ and $D$ waves, with a branching ratio of 
95\% \cite{PDG}. To a good approximation one can neglect the $D$ wave
component, which allows a parameter-free computation of the asymmetry.
The smaller $D$-wave component and the $K\rho$ contribution can be
also included using the measurements of the partial wave amplitudes 
and phases from the ACCMOR Collaboration \cite{ACCMOR}.

The most convenient way of presenting the result for the polarization
sensitive observable is in terms of an angular distribution in the 
rest frame of the resonance $K_{\rm res}$. Introducing the angle
$\theta$ between the opposite of the photon momentum $-\vec q$ and
the normal to the $K\pi\pi$ decay plane defined as $\vec p_{\rm slow}\times
\vec p_{\rm fast}$, where $\vec p_{\rm slow}$ and $\vec p_{\rm fast}$
are the momenta of the slower and faster pions, this is given by \cite{GroPi}
\bea\label{3res}
& & \frac{\mbox{d}^2\Gamma}{\mbox{d}s \mbox{d}\cos\tilde\theta} = 
|c_1|^2  \left\{
1 + \cos^2\tilde \theta + 4 P_\gamma R_1 \cos\tilde \theta \right\}\\
&+&  |c_2|^2  \left\{
\cos^2\tilde\theta + \cos^2 2\tilde\theta + 
12 P_\gamma R_2 \cos\tilde\theta \cos 2\tilde\theta\right\}
 + |c_3|^2 B_{K^*_1}(s) \sin^2\tilde\theta\nonumber\\
&+& 
\left\{ c_{12} \frac12(3\cos^2\tilde\theta -1) + 
P_\gamma c'_{12}
 \cos^3\tilde\theta \right\}\nonumber\,,
\eea
where the first three terms are produced by decays through $K_{\rm res}$
resonances with $J^P = 1^+, 2^+$ and $1^-$, and the last terms come from
$1^+-2^+$ interference, respectively.
The hadronic parameters $R_{1,2}$ can be computed with relatively small
model dependence as explained above, which gives \cite{GGPR,GroPi}
$R_1 = 0.22 \pm 0.03$, $R_2 = 0.01 - 0.05$. Thus, measurements of the
angular distribution (\ref{3res}) can be used to extract the photon
polarization parameter $P_\gamma$. 

Assuming dominance of the $J^P = 1^+$ resonance $K_1(1400)$, the
angular distribution (\ref{3res}) predicts an up-down asymmetry 
of the photon momentum direction relative to the normal to the $K\pi\pi$
plane $A_{\rm up-down} = \frac32 R_1 P_\gamma$. The significant value
of this asymmetry makes this channel particularly attractive.

Assuming an exclusive branching ratio 
$Br(B\to K_1(1400)\gamma) = 0.7\times 10^{-5}$ and taking the
final state in (\ref{chain}) to be detected through the $K^+\pi^-\pi^0$
and $K_S\pi^+\pi^0$ modes, implies that about $2\times 10^7$ $B\bar B$ pairs
are required to measure 80 $K\pi\pi\gamma$ events which should be
sufficient for a $3\sigma$ confirmation of a left-handed photon in 
$b\to s\gamma$ decay. Such a measurement should be feasible at the existing
$B$ factories in the near future.

\bigskip
It is a pleasure to thank Michael Gronau, Yuval Grossman and Anders Ryd for
an enjoyable collaboration. I am grateful to J\"org Urban for comments on the
manuscript. This work has been supported by the DOE under Grant 
No. DOE-FG03-97ER40546.

\end{document}

%% file: econfmacros.tex
\def\beq{\begin{equation}}
\def\eeq#1{\label{#1}\end{equation}}
\def\eeqn{\end{equation}}

\def\beqa{\begin{eqnarray}}
\def\eeqa#1{\label{#1}\end{eqnarray}}
\def\eeqan{\end{eqnarray}}

\let\bar=\overbar

\def\Dslash{\not{\hbox{\kern-4pt $D$}}}
\def\dslash{\not{\hbox{\kern-2pt $\del$}}}

\def\msb{{\bar{\ssstyle M \kern -1pt S}}}

